\documentclass[12pt,preprint]{aastex}      

\usepackage[latin1]{inputenc}
\usepackage{lscape}
\usepackage{amsmath}

\slugcomment{To appear in ApJL}

\shorttitle{Imaging the Gaps in the Disk of HD 169142 at 7 mm}

\shortauthors{Osorio et al.}

\begin{document}

 \title{Imaging the Inner and Outer Gaps of the Pre-Transitional Disk 
of HD 169142 at 7 mm
}

\author{Mayra Osorio\altaffilmark{1}, 
Guillem Anglada\altaffilmark{1},
Carlos Carrasco-Gonz\'alez\altaffilmark{2},
Jos\'e M. Torrelles\altaffilmark{3},
Enrique Mac\'{\i}as\altaffilmark{1},
Luis F. Rodr\'{\i}guez\altaffilmark{2},
Jos\'e  F. G\'omez\altaffilmark{1},
Paola D'Alessio\altaffilmark{2},
Nuria Calvet\altaffilmark{4},
Erick Nagel\altaffilmark{5},
William R. F. Dent\altaffilmark{6},
Sascha P. Quanz\altaffilmark{7},
Maddalena Reggiani\altaffilmark{7},
Juan M. Mayen-Gijon\altaffilmark{1}
}

\altaffiltext{1}{Instituto de Astrof\'\i sica de Andaluc\'\i a (CSIC), 
Glorieta de la Astronom\'\i a s/n, E-18008 Granada, Spain;  email: 
{\tt osorio@iaa.es}}
 \altaffiltext{2}{Centro de Radioastronom\'{\i}a y Astrof\'{\i}sica 
UNAM, Apartado Postal 3-72 (Xangari), 58089 Morelia, Michoac\'an, Mexico}
\altaffiltext{3}{Institut de Ci\`encies de l'Espai (CSIC)-Institut de 
Ci\`encies del Cosmos (UB)/IEEC, Mart\'{\i} i Franqu\`es 1, E-08028 
Barcelona, Spain}
 \altaffiltext{4}{Department of Astronomy, University of Michigan,  825 
Dennison Building, 500 Church St, Ann Arbor, MI 48109, USA}
 \altaffiltext{5}{Departamento de Astronom\'{\i}a, Universidad de 
Guanajuato, Guanajuato, Gto 36240, Mexico}
\altaffiltext{6}{ALMA SCO, Alonso de C\'ordova 3107, Vitacura, Santiago, 
Chile}
 \altaffiltext{7}{Institute for Astronomy, ETH Zurich, 
Wolfgang-Pauli-Strasse 27, 8093 Zurich, Switzerland}


\begin{abstract}

We present Very Large Array observations at 7 mm that trace the thermal 
emission of large dust grains in the HD 169142 protoplanetary disk. Our images
 show a ring of enhanced emission of radius $\sim$25-30 AU, whose inner region 
is devoid of detectable 7 mm emission. We interpret this ring as tracing the 
rim of an inner cavity or gap, possibly created by a planet or a substellar 
companion. The ring appears asymmetric, with the western part significantly 
brighter than the eastern one. This azimuthal asymmetry is reminiscent of the 
lopsided structures that are expected to be produced as a consequence of 
trapping of large dust grains. Our observations also reveal an outer annular 
gap at radii from $\sim$40 to $\sim$70 AU. Unlike other sources, the radii of 
the inner cavity, the ring, and the outer gap observed in the 7 mm images, 
which trace preferentially the distribution of large (mm/cm sized) dust grains, 
coincide with those obtained from a previous near-infrared polarimetric image, 
which traces scattered light from small (micron-sized) dust grains. We model 
the broad-band spectral energy distribution and the 7 mm images to constrain 
the disk physical structure. From this modeling we infer the presence of a 
small (radius $\sim$0.6 AU) residual disk inside the central cavity, indicating 
that the HD 169142 disk is a pre-transitional disk. The distribution of dust in 
three annuli with gaps in between them suggests that the disk in HD 169142 is 
being disrupted by at least two planets or substellar objects.

\end{abstract}

\keywords{planet-disk interactions --- protoplanetary disks --- stars: 
formation --- stars: individual (HD 169142) --- stars: pre-main 
sequence}

\section{Introduction}

The early stages of planet formation are expected to be identified by 
signatures of dust evolution in the protoplanetary disks. An outstanding 
signature is the large (several tens AU in diameter) central cavities 
almost devoid of dust that characterize the so-called transitional disks 
(Calvet et al. 2005). Some of these transitional disks, dubbed 
pre-transitional disks, show an important near-IR excess that is 
interpreted as produced by a residual inner disk inside the cavity 
(Espaillat et al. 2008).

A number of processes have been proposed to explain the origin of these 
cavities. Photoevaporation and winds can remove material from the inner 
disk (e.g., Alexander et al. 2006; Suzuki et al. 2010); grain-growth can 
diminish the dust opacity, decreasing its continuum emission (Dullemond 
\& Dominik 2005); finally, dynamical clearing due to tidal interactions 
with low-mass companions, either very young brown dwarfs or giant 
planets on long-period orbits, can produce annular gaps and cavities in 
the disk (Papaloizou et al. 2007; Zhu et al. 2011). Andrews et al. 
(2011) favored the last of these mechanisms in their analysis of a 
sample of 12 such disks. Therefore, transitional and pre-transitional 
disks are excellent places to search for the youngest planets.

The existence of central cavities (first suggested by Strom et al. 1989) 
was confirmed from the modeling of the spectral energy distributions 
(SEDs) and the analysis of Spitzer spectra (e.g., D'Alessio et al. 
2006). However, in recent years, it has been possible to image several 
of these cavities through mm and submm interferometric observations 
(Brown et al. 2009; Andrews et al. 2011; Isella et al. 2013; Casassus et 
al.  2013), as well as by high angular resolution polarimetric imaging 
at infrared wavelengths (e.g., Quanz et al. 2011; Hashimoto et al. 2012; 
Garufi et al. 2013; Avenhaus et al. 2014).

The disk surrounding the Herbig Ae/Be star HD 169142 
($M_*\simeq2~M_{\odot}$, $d=145$ pc; Manoj et al. 2006 and references 
therein) is one of the best targets to study the planetary formation 
process. Near-infrared polarization images as well as millimeter 
continuum and CO observations reveal that the disk is nearly face-on, 
extending up to $\sim$240 AU (e.g., Kuhn et al. 2001; Raman et al. 
2006). Its SED characteristics and its relatively strong 7 mm emission 
suggest that dust has evolved to form big grains (Dent et al. 2006), a 
prerequisite to planet formation.

Modeling of the SED of HD 169142 suggested the presence of a disk inner 
cavity with an estimated radius of $\sim$40 AU (Grady et al. 2007) or 
$\sim$20 AU (Meeus et al. 2010; Honda et al. 2012; Maaskant et al. 
2013). These results are supported by recent H-band polarized light 
images obtained with NACO/VLT (Quanz et al. 2013) that reveal a bright 
emission ring with a radius of $\sim$25 AU, which is interpreted as the 
rim of the inner cavity. Additionally, these images reveal the presence 
of a $\sim$40-70 AU annular gap. However, with these data alone it 
cannot be unambiguously established whether this gap in polarized 
emission corresponds to a real gap in surface density, possibly induced 
by one (or several) forming planet(s), or if it is an illumination 
effect.

We present Very Large Array (VLA) observations towards HD 169142 that 
reveal, for the first time, the structure of the dust thermal emission 
of the disk at radii from $\sim$10 to $\sim$100 AU.

\section{Observations}

Observations were carried out with the Karl G. Jansky VLA of the 
National Radio Astronomy Observatory (NRAO)\footnote{The NRAO is a 
facility of the National Science Foundation operated under cooperative 
agreement by Associated Universities, Inc.}. We observed the continuum 
emission in the Q ($\sim$7 mm), C ($\sim$5.5 cm) and X ($\sim$3.3 cm) 
bands in three different array configurations (CnB, A, and B) (see Table 
1).
 Calibration of the complex gain of the antennas was performed by 
observing J1820-2528 every 2 minutes.  We estimate that the uncertainty 
in the absolute positions in Q band is $\sim 0.1''$. The flux scale was 
set by observing 3C286.

Data were calibrated with the CASA (Common Astronomy Software 
Applications; version 
4.1.0)\footnote{https://science.nrao.edu/facilities/vla/data-processing} 
package. For each data set, we ran the VLA Calibration Pipeline provided 
by the NRAO (version 
1.2.0)\footnote{https://science.nrao.edu/facilities/vla/data-processing/pipeline}. 
After inspection of the calibrated data, we performed additional data 
flagging and re-ran the pipeline when needed.

We produced deconvolved images from each configuration data set, as well 
as an image combining data from all configurations. All images were 
obtained with the multi-scale, multi-frequency CLEAN algorithm (Rau \& 
Cornwell 2011) and implemented in the task {\em clean} of CASA.

\section{Results} 

We did not detect any significant emission at C ($\sim$5.5 cm) or at X 
($\sim$3.3 cm) bands, with 3-$\sigma$ upper limits of 36 and 39 $\mu$Jy 
beam$^{-1}$, respectively (Table 1). By extrapolating the X-band upper 
limit with a spectral index of +0.6, a value typical of free-free 
emission from thermal radio jets (e.g., Anglada 1996), we expect a 
free-free contribution at 7 mm of $<0.09$ mJy, $<5\%$ of the observed 
flux density. Thus, the 7 mm emission reliably traces the thermal dust 
emission of the disk with negligible free-free contamination.

In Figure 1 we show VLA images at 7 mm wavelength of the disk 
surrounding HD 169142.  The source is marginally resolved in the CnB 
configuration image (Fig. 1a), showing that the emission is asymmetric, 
with the western side being brighter than the eastern side. The source, 
with a total flux density of 1.8$\pm$0.3 mJy, is well resolved in the 
sensitive (8 GHz bandwidth) B configuration image (Fig. 1b) showing a 
ring of emission of radius $\sim0.18''$-$0.20''$ ($\sim$ 25-30 AU). The 
width of the ring appears unresolved, even in the E-W direction where 
the angular resolution is better. The decrease in intensity observed 
towards the northern and southern edges of the ring is an observational 
effect, since in these directions the north-south elongated beam 
intersects a smaller fraction of the ring area.

The center of the emission ring in the B configuration image is located 
at $\alpha$(J2000) = 18$^{\rm h}$24$^{\rm m}$29.776$^{\rm s}$, 
$\delta$(J2000) = $-$29$^\circ$46$'$49.87$''$, which falls 0.04$''$ west 
and 0.03$''$ north of the nominal position of the star HD 169142 in the 
Tycho-2 catalog, after correction for proper motions to the epoch of the 
observation. Therefore, the VLA observations indirectly provide an 
accurate position for the star.
 In the A configuration images, the emission ring is only marginally 
detected, after setting a maximum uvrange of $\sim$1000 k$\lambda$. An 
image obtained from the CnB and A configuration data is presented in 
Osorio et al. (2014).

The image in Figure 1c was obtained by combining the interferometric 
visibilities from the CnB, B, and A configurations. To correct for 
residual differences in absolute positions due to phase errors and 
proper motions, the data from the A and B configurations were aligned 
(within $<$0.05$''$) by matching the centers of the rings fitted to the 
individual images before being combined. For the CnB data, obtained with 
a larger beam, a precise alignment was considered unnecessary. The final 
image shows the same general structure as the one seen with the B-array 
data alone, but with better detail.

\subsection{The Emission Ring and the Inner Cavity/Gap}

One of the outstanding features of the HD 169142 disk images is the ring 
of enhanced emission of radius $\sim$25-30 AU. We interpret this ring as 
the rim of an inner cavity or gap possibly created in the disk by a 
planet or a substellar companion. Figure 1d shows an overlay of the VLA 
image from Figure 1c and the H-band polarized light image of 
Quanz et al. (2013). The cavity is suggested by the decrease of emission 
near the inner edge of the ring in the IR image, but it cannot be traced 
towards the center because the central pixels are saturated. 
Nevertheless, this inner cavity is clearly seen in the 7 mm images 
(Figs. 1b, c, d), confirming HD 169142 as a transitional or 
pre-transitional disk. Interestingly, the size of the cavity in the IR 
image, which traces scattered light from small (micron-sized) dust 
grains, coincides with that in the 7 mm image, which traces 
larger (mm/cm-sized) dust grains. This is in contrast with other 
sources where the size of the cavity changes with the observed 
wavelength (see Garufi et al. 2013 and references therein).

The 7 mm emission ring appears significantly asymmetric (in the 
B-configuration image, the flux densities of the western and eastern 
halves of the ring are 630$\pm$105 mJy and 361$\pm$78 mJy, respectively, 
without overlapping of the quoted 99\% confidence intervals). This 
azimuthal asymmetry is reminiscent of the lopsided rings that are 
expected to be produced at long wavelengths as a consequence of 
azimuthal accumulation (trapping) of large dust grains, as predicted by 
theoretical models (e.g., Birnstiel et al. 2013) and revealed by ALMA 
observations of more extreme asymmetries in sources such as Oph IRS 48 
(van der Marel et al. 2013) and HD 142527 (Casassus et al. 2013). In 
near-IR scattered light the emission ring of HD 169142 is more symmetric 
but shows a dip at PA $\simeq80^\circ$ (Quanz et al. 2013), and perhaps 
a second dip at PA $\simeq150^\circ$ (Fig. 1c), both in the eastern 
side, where the 7 mm emission is weaker.

\subsection{The Outer Annular Gap}

Quanz et al. (2013) reported the presence of an annular gap in the IR 
polarized light of the disk of HD 169142 in the range of radii 
$\sim$40-70 AU.
 This gap in scattered light may be due to a real decrease in the disk 
surface density, which could be induced by a protoplanet, or may be due 
to shadowing or other illumination effect that decreases the polarized 
light. Our 7 mm observations, tracing optically thin thermal emission of 
the large dust grains, could provide a better determination of the real 
nature of this gap. However, the 7 mm emission is weak at $\ga$40 AU 
from the center. To improve the signal to noise ratio at large radii, we 
have calculated the radial intensity profile, shown in Figure 2a, 
averaging the emission over concentric rings in Figure 1c.
 To highlight the outer gap and its radial location, Figure 2b shows the 
same profile but normalized by a power-law fit to compensate for the 
radial intensity decrease. In the azimuthal averaging we have excluded 
PAs near $\sim180^\circ$, where there is a knot of emission that may be 
an independent component (see below).  The radial intensity profile 
shows that the outer gap, as well as the inner cavity and the bright 
ring, are found at the same radii as the corresponding features in the 
IR profile (Fig. 2 in Quanz et al. 2013), suggesting that they represent 
the true dust distribution.

An additional interesting feature observed in our 7 mm VLA images is the 
knot of emission located $\sim0.34''$ ($\sim$50 AU) to the south (PA 
$\simeq175^\circ$) of the central star. This emission appears smeared 
out in the B configuration image (Fig. 1b) but is better defined in the 
image that combines all configurations (Fig. 1c).
  This compact source falls in the middle of the outer annular gap, and 
we speculate that it could trace circumplanetary dust emission 
associated with the protoplanet responsible for creating this gap. We 
estimate for this knot a flux density of $\sim$100 $\mu$Jy above the 
background ($\sim$5-$\sigma$). Assuming a temperature of $\sim$50 K and 
a 7 mm opacity of 2$\times10^{-3}$ cm$^{-2}$ g$^{-1}$, as suggested by 
the SED modeling (section 4), we obtain a total (dust+gas) mass of 0.6 
$M_J$. This mass estimate is uncertain, since in the proximity of gaps 
the size of dust grains and the dust-to-gas ratio are expected to change 
with time and planet mass (Pinilla et al. 2012). If the 7 mm source is 
associated with a protoplanet, its expected orbital period ($\sim$186 
yr) would produce proper motions detectable in a few years, providing an 
unambiguous way to test this hypothesis.

\section{Disk Model}

HD 169142 has been observed from ultraviolet to radio wavelengths. 
Figure 3a shows the photometric data points compiled from the literature 
and from this paper. Our model includes the contribution of the central 
star (whose adopted parameters are given in Table 2) and the disk. As 
suggested by the observations (Quanz et al. 2013; this paper) the disk 
has a central cavity of radius $\sim$30 AU and a gap spanning 
$\sim$40-70 AU. We find that a central hot component is required to fit 
the 2-10 $\mu$m range of the SED. A central dust component was also 
inferred from the modeling of Honda et al. (2012) and Maaskant et al. 
(2013), who postulate the presence of a central dust halo. However, we 
find that the emission can be naturally explained by a small residual 
inner disk and its hot wall (located where dust reaches its sublimation 
temperature). With this residual inner dust, the disk of HD 169142 
should be classified as pre-transitional (as defined by Espaillat et al. 
2008), implying that the observed central cavity is actually a gap 
between the inner and outer parts of the disk (see sketch in Fig. 2c).

The contribution of the disk is obtained using the irradiated 
$\alpha$-accretion disk models with dust settling developed by D'Alessio 
et al. (1999, 2001, 2006). In these models, the radial and vertical 
physical structure of the disk, and the emerging intensity are 
self-consistently calculated. Since HD 169142 is a very well studied 
object, most of the parameters of the model are determined by previous 
observations; other are pretty well constrained, and only small 
fine-tuning adjustments were required (see Table 2 and references 
therein).  The main free parameters of the model are the viscosity and 
the dust properties. A low value of the viscosity parameter is required 
to account for both the low mass accretion rate and the high surface 
densities (suggested by the relatively strong mm/submm flux densities), 
as in $\alpha$-disks the viscosity parameter is proportional to the mass 
accretion rate and inversely proportional to the surface density.

The dust is assumed to consist of two populations of grains (each with 
an $n(a)\propto a^{-3.5}$ size distribution) that are vertically 
distributed as a function of the degree of settling of large grains from 
the upper layers. The grain sizes are constrained mainly by the data at 
wavelengths $>$100 $\mu$m. Grains in the disk upper layers have radii 
ranging from 0.005 $\mu$m to 1 $\mu$m, while the radii of the grains 
settled in the disk mid-plane range from 5 $\mu$m to 1 mm. The 
dust-grain mixture assumed to compute the opacity consists of silicates 
and graphite, with mass fractional abundances relative to gas of 0.004 
and 0.0025, respectively (e.g., McClure et al. 2013).

The inner disk sublimation wall has a curved shape that depends on 
density and grain settling, and is calculated following Nagel et al. 
(2013) and D'Alessio et al. (2006). For the gaps and their walls we use 
a simpler treatment (see, e.g., Jang-Condell \& Turner 2012 for a more 
accurate treatment). The gaps are modeled as annular regions completely 
devoid of dust with cylindrical walls whose scale height is predicted by 
the accretion disk model. The wall radii are obtained approximately from 
the observed images (Quanz et al. 2013; this paper) and refined (within 
$\sim\pm1$ AU) by fitting the SED and the 7 mm intensity profile. 

The emission of the outer walls of the gaps is computed as a modified 
blackbody (D'Alessio et al. 2005), but taking into account the 
extinction by the disk. Thus, at short wavelengths, just half of the 
inner face of the wall will be visible, while at 7 mm we will be 
detecting the whole wall. Walls are assumed to be unresolved in the 
radial direction, with a width smaller than the pixel scale ($\sim$1 
AU). Because of shadowing, only the upper portion (composed of small 
dust grains) of these walls is frontally irradiated by the star, and is 
heated at a temperature close to the equilibrium temperature, which is 
calculated following Nagel et al. (2013). The contribution of the 
shadowed portion of the walls (at lower temperatures, defined by the 
disk structure) is relevant in the 7 mm intensity profile. This emission 
is dominated by the large dust grains (close to the disk mid-plane) that 
accumulate near the outer edges of the gaps because of dust filtration. 
To simulate this effect, in these walls we increased the dust-to-gas 
ratio by a factor of $\sim$3, and the scale height of the large dust 
grain population up to 60\% of the local gas-pressure scale height.
These two parameters are determined mainly by fitting the 7 mm 
intensity profile.

The main parameters of our favored model are given in Table 2. Figure 3a 
shows that the model reproduces the observed SED satisfactorily. Figure 
3b shows a CASA simulated image of the 7 mm model emission that agrees 
in shape, size, and intensity with the observed image (Fig. 1c). Figures 
2a and 2b show a quantitative comparison of the radial intensity 
profiles of both images, showing that model and observations are in good 
agreement. The contribution of the hot inner disk and its wall (dominant 
in the 2-10~$\mu$m range), remain below the detectability limits at 7 
mm, as expected. Very sensitive high angular resolution observations at 
mm/submm wavelengths are necessary to directly confirm the presence of 
this disk component.

The simulated model image (Fig. 3b) appears symmetric with respect to 
the beam axis. Subtraction from the observed one shows a residual 
emission excess in the western side, indicating that the east-west 
asymmetry in the observed images (Fig. 1) cannot be attributed to 
opacity effects and should have a different origin, such as a dust trap.

As noted above, most of the physical structure of the HD 169142 disk is 
self-consistently modeled and constrained by the observations. However, 
the gaps and walls arising from the tidal interactions of the disk with 
companions, and the subsequent radial and azimuthal gradients of the 
gas-to-dust ratio, are not self-consistently integrated. Model 
improvements should come from including azimuthal asymmetries and dust 
migration.

In summary, our observational and model results show that the disk of HD 
169142 is a pre-transitional disk with two gaps in the dust 
distribution, suggesting that it is being disrupted by at least two 
planets or substellar objects. Actually, Reggiani et al. (2014) have 
just reported 3.8 $\mu$m observations revealing the presence of a 
low-mass companion in the inner gap of the disk, at a separation of 
$\sim$23 AU.

\acknowledgments

We thank an anonymous referee for his/her valuable comments. G.A., 
C.C.-G., E.M., J.F.G., J.M.M.-G., M.O., and J.M.T. acknowledge support 
from MICINN (Spain) grant AYA2011-30228-C03 (co-funded with FEDER funds) 
and from Junta de Andaluc\'{\i}a (TIC-126).  C.C.-G., P.D., and L.F.R. 
acknowledge the support of DGAPA, UNAM, and CONACyT (Mexico).

{\it Facilities:} \facility{VLA}

\clearpage



\begin{figure}
\plotone{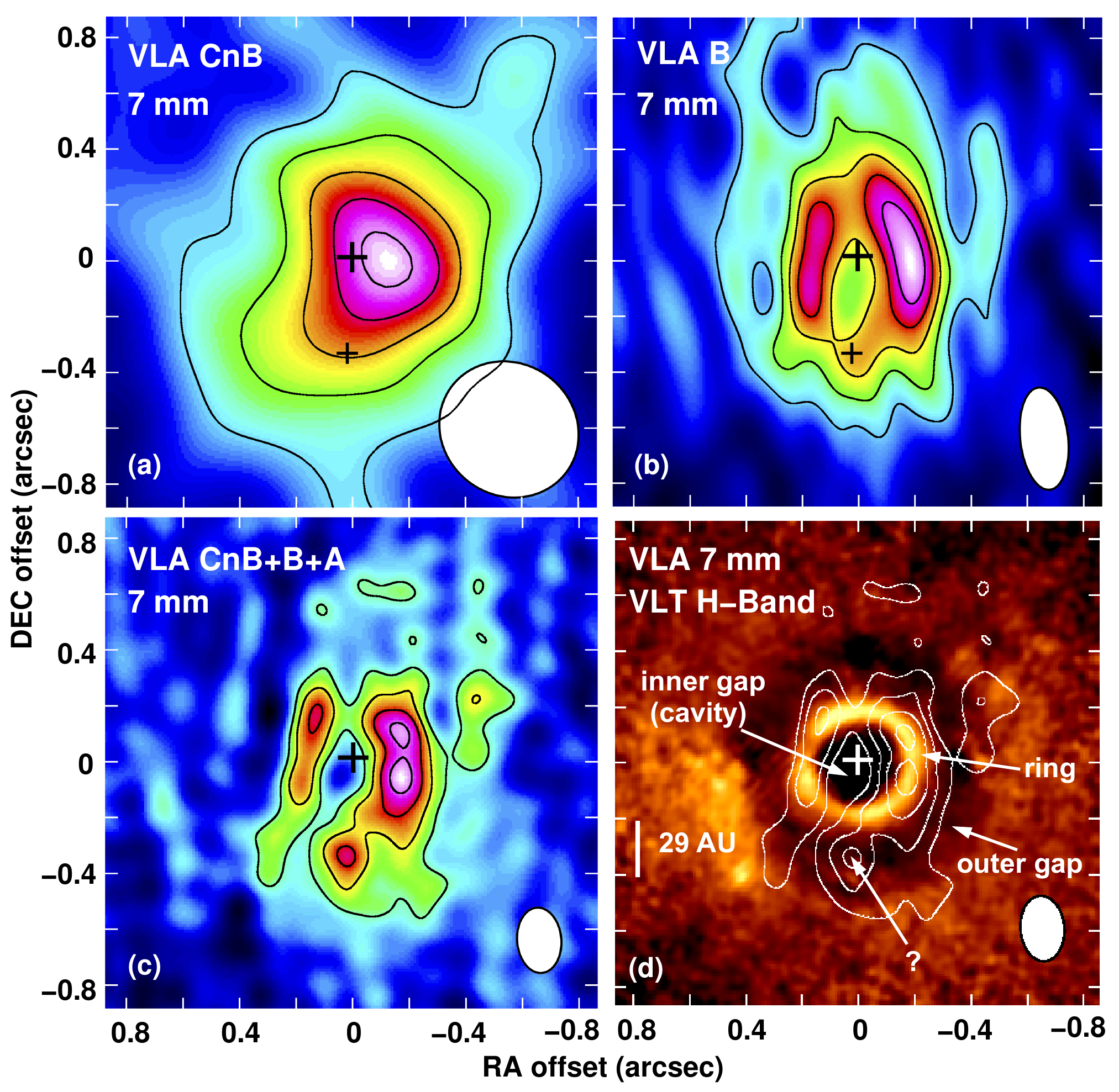}
 \caption{VLA images of the 7 mm dust thermal emission in several array 
configurations. Panels {\em (a)} and {\em (b)} show, respectively, the 
CnB and B configuration images. Panel {\em (c)} shows the image obtained 
by combining the CnB, B, and A configuration visibilities with a 
uvrange $<$1500 k$\lambda$ (rms=18 $\mu$Jy beam$^{-1}$; 
beam=$0.23''\times0.16''$, PA=$5^\circ$). Panel {\em (d)} shows an 
overlay of the image shown in panel {\em (c)} (contours) and the 
VLT/NACO H-band (1.6 $\mu$m) polarized light image from Quanz et al. 
(2013) (color-scale). Saturated pixels in the central region of the 
H-band image have been masked out. In all panels, contour levels are 
$-$3, 3, 5, 7, 9, and 11 times the rms. Synthesized beams are 
plotted in the lower-right corners. The apparent decrease of the 7 mm 
emission in the north and south edges of the source is most probably 
a consequence of the elongated beam. The larger cross marks the position 
of the HD 169142 star and the smaller one that of the protoplanet 
candidate.}
 \label{Fig1}
 \end{figure}

\clearpage

\begin{figure}
\epsscale{0.8}
\plotone{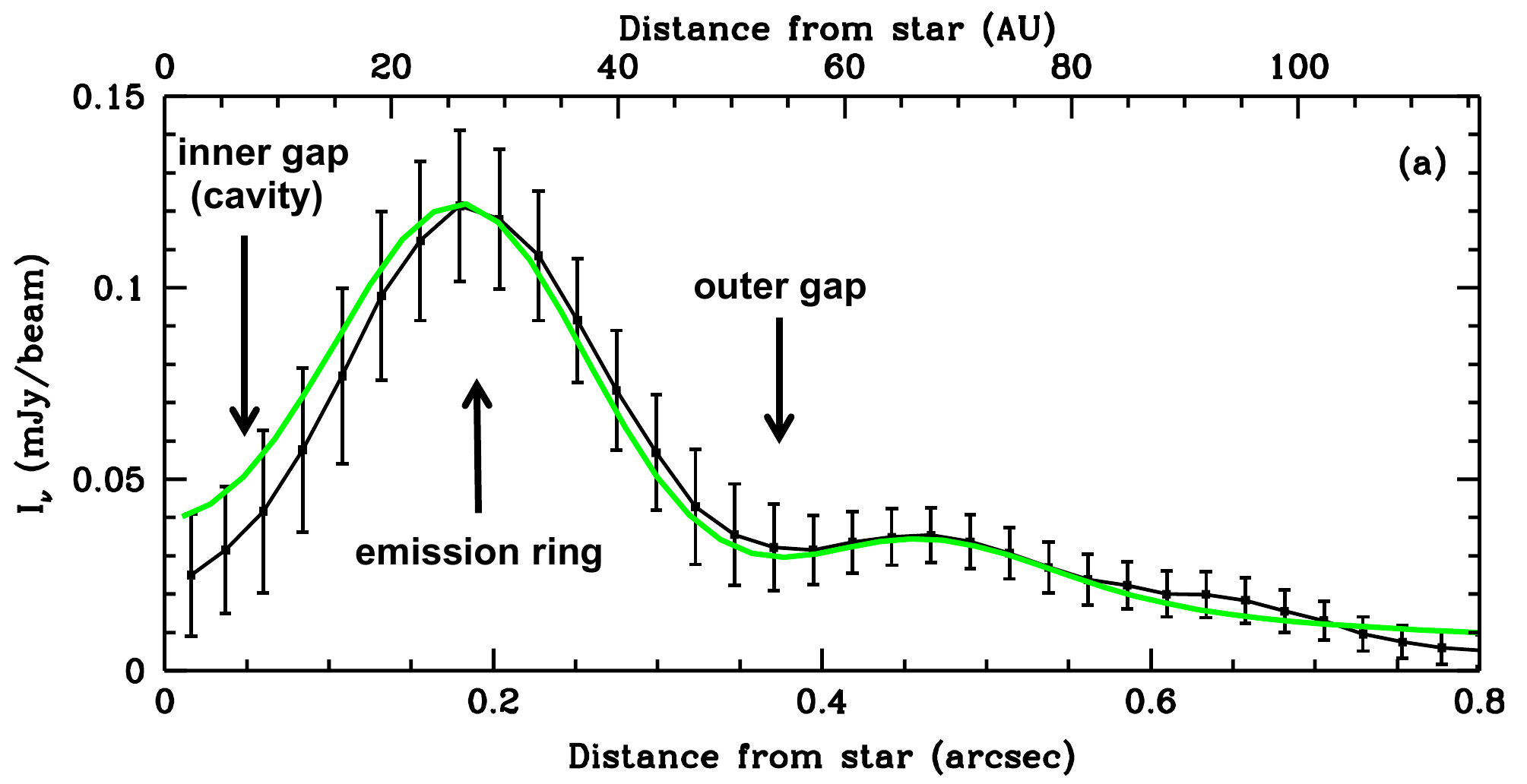}
\vspace{0.2cm}
\plotone{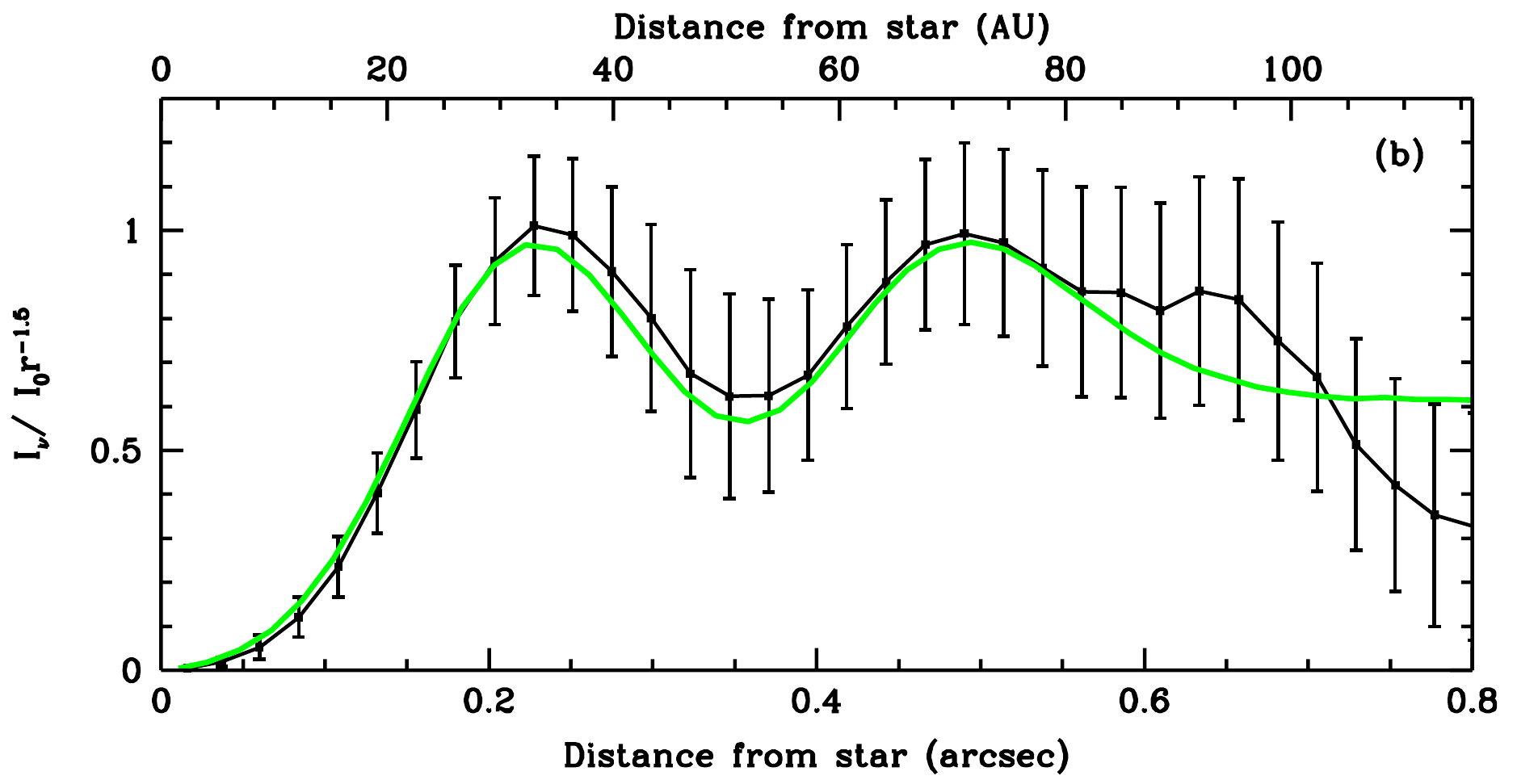}
\vspace{0.2cm}
\plotone{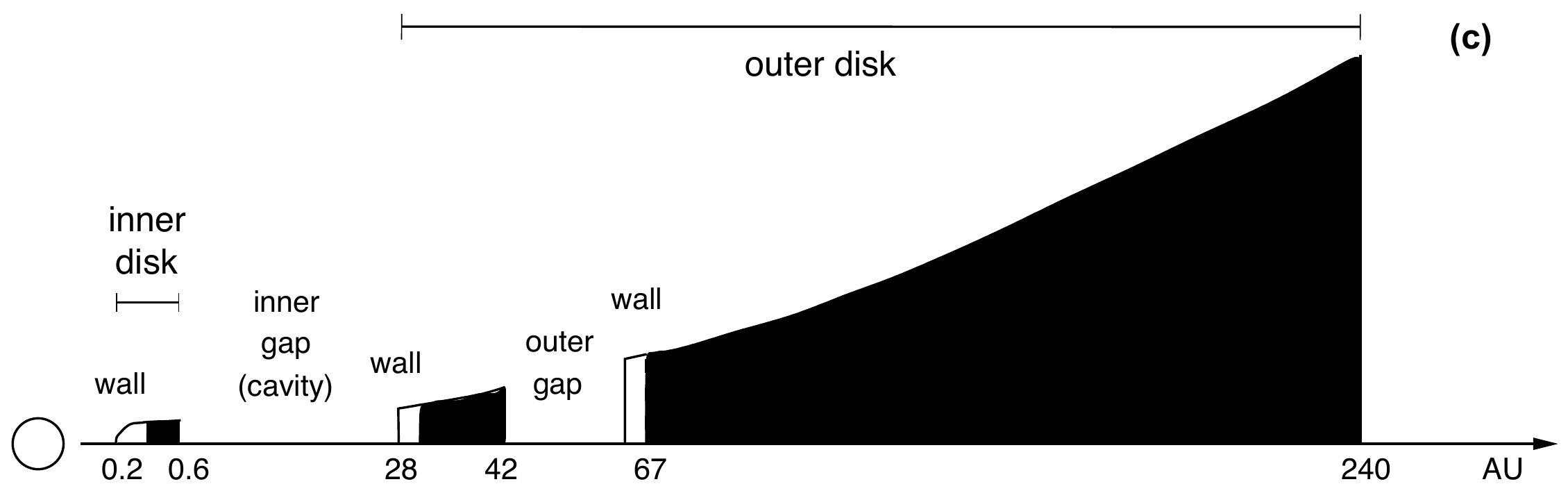}
 \caption{(a) Observed (dots and black solid line with error bars) and 
model (green solid line) azimuthally averaged intensity profile of the 
images shown in Fig. 1c and in Fig. 3b, respectively. Error bars in the 
observed profile are obtained from the rms of the pixels in the annuli 
corrected by the number of uncorrelated pixels. (b) Same as (a), but 
normalized by a power-law fit. (c) Sketch of the disk structure above 
the mid-plane.}
 \epsscale{1.0}
 \label{Fig2}
 \end{figure}

\clearpage

\begin{figure}
\epsscale{.55}
\plotone{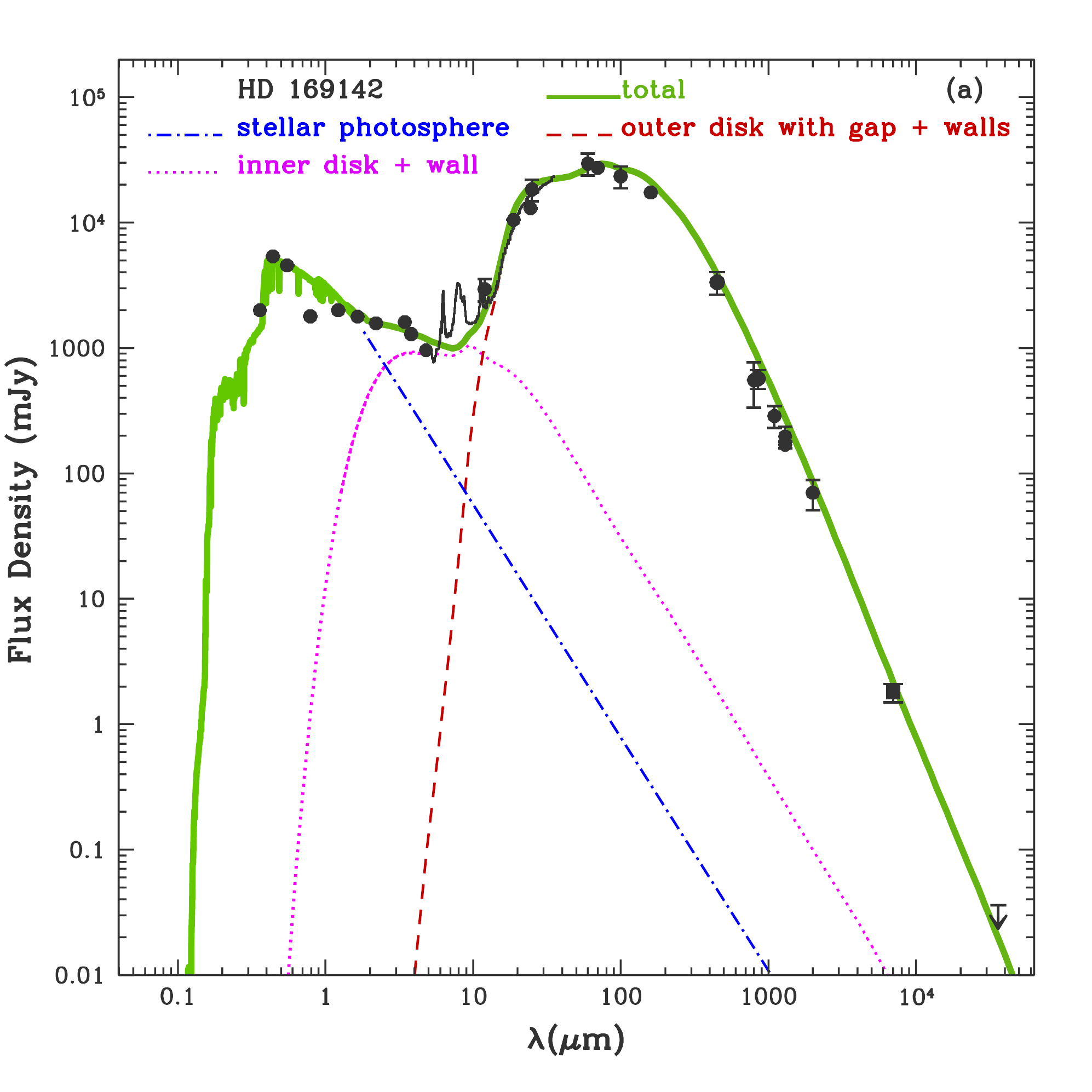}
\plotone{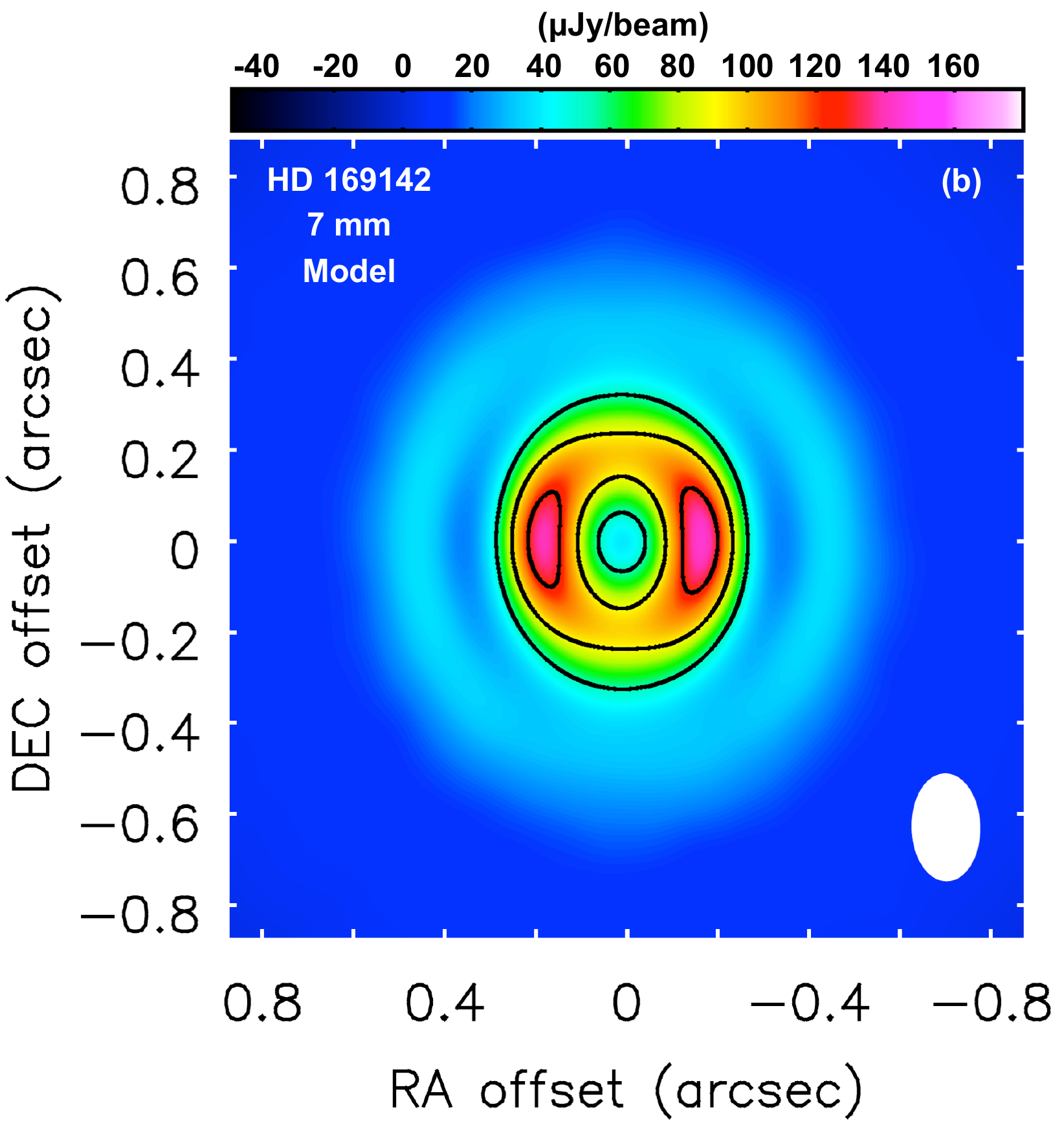}
 \caption{(a) Observed (points) and model (lines) SEDs of the HD 
169142 protoplanetary disk. The thick solid line is the total SED, 
including the contribution of all the components. Data points are from 
the compilation of Dent et al. 2006, from Raman et al. 2006, Honda et 
al. 2012, Meeus et al. 2010, Sandell et al. 2011, Dempsey et al. 2013, 
and this paper.  (b) CASA simulated image at 7 mm for the disk 
model, as it would be observed by combining the VLA CnB, B, and A 
configurations. Contours and color scale are as in Fig. 1c.}
 \label{Fig3}
 \end{figure}

\clearpage


\begin{deluxetable}{lcccccccc}
\tabletypesize{\scriptsize}
\tablewidth{0pt}
\tablecaption{Summary of VLA Observations\tablenotemark{a}}
\tablehead{
\colhead{Observation} & 
\colhead{Project} & 
\colhead{Array} & 
&
\colhead{Frequency} & 
\colhead{Mean} &
\colhead{rms} & 
\multicolumn{2}{c}{Synthesized Beam\tablenotemark{b}}\\
\cline{8-9} 
\colhead{Date} &
\colhead{Code} &
\colhead{Configuration} &
\colhead{Band} & 
\colhead{Range} &
\colhead{Wavelength} &
\colhead{Noise\tablenotemark{b}} &
\colhead{Size} &
\colhead{PA} \\
& & & &
\colhead{(GHz)} &
\colhead{(mm)} &
\colhead{($\mu$Jy~beam$^{-1}$)} &
\colhead{(arcsec$\times$arcsec)} &
\colhead{(deg)}
}
\startdata
2012-May-03 & 12A-439 &  CnB  &   Q  &  42-44  & 7.0  & 28  & 0.51$\times$0.47 & 50\\
2012-May-03 & 12A-439 &  CnB  &   C  &  4.5-6.5  & 55 & 12  & 4.28$\times$3.76 & 48 \\
2012-Nov-12 & 12A-439 &   A   &   Q  &  42-44  & 7.0  & 25  & 0.11$\times$0.05 & $-$4 \\
2012-Nov-12 & 12A-439 &   A   &   C  &  4.5-6.5  & 55 & 12 & 0.85$\times$0.39 & $-$1 \\
2013-Sep-28 & 13B-260 &   B   &   Q  &  40-48  & 6.8  & 17 & 0.37$\times$0.17 & 6 \\
2013-Sep-28 & 13B-260 &   B   &   X  &   8-10  & 33 & 13  & 1.76$\times$0.77 & 8 \\
\enddata
\tablenotetext{a}{Phase center was at 
$\alpha$(J2000)=18$^{\rm h}$24$^{\rm m}$29.7790$^{\rm s}$, 
$\delta$(J2000)=$-$29$^\circ$46$'$49.547$''$.}
\tablenotetext{b}{For naturally weighted maps.}
\end{deluxetable}

\begin{deluxetable}{lccc}
\tablewidth{0pt}
\tablecaption{Physical Parameters of HD 169142 and its Disk\label{Tab2}}
\tablehead{
\colhead{Parameter} &
\colhead{Value} &
\colhead{Notes} &
\colhead{Refs.}
}
\startdata
\multicolumn{4}{c}{Stellar Properties} \\
\hline
Distance (pc)  & 145 & Adopted & 1\\
$A_V$ (mag) & 0.5 & Adopted & 2\\
Age (Myr) & 5.4 & Adopted & 1\\
Effective Temperature (K) & 8100 & Adopted & 1\\
Radius ($R_{\odot})$  & 2.2 & Adopted &1\\
Mass  ($M_{\odot}$)   & 2 & Adopted & 1\\
\hline
\multicolumn{4}{c}{Inner Disk\tablenotemark{a}} \\
\hline
Inner Radius (AU) & 0.5 & Fitted \\
Outer Radius (AU)  & 0.6 & Fitted \\
Mass (M$_{\odot}$) & 2.6$\times$10$^{-5}$ & Calculated\\
\hline
\multicolumn{4}{c}{Wall of the Inner Disk} \\
\hline
Location (AU) & 0.2-0.5 & Calculated \\
Temperature (K) & 1300-1000 & Calculated \\
Height (AU) & 0-0.05 & Calculated \\
\hline
\multicolumn{4}{c}{Outer Disk} \\
\hline
Inclination Angle (deg) & 13 & Adopted & 3 \\
Position Angle (deg) & 5 & Adopted & 3 \\
Inner Radius (AU)  & 28 & Adopted/Refined & 4, 5  \\
Outer Radius (AU)  & 240 & Adopted & 3 \\
Gap Inner Radius (AU)  & 42 & Adopted/Refined & 4, 5  \\
Gap Outer Radius (AU)  & 67 & Adopted/Refined & 4, 5 \\
Mass Accretion Rate ($M_{\odot}$ yr$^{-1}$) & 3$\times$10$^{-9}$ &
Adopted/Refined & 6 \\
Viscosity Coefficient & 0.0005 & Fitted\\
Degree of Settling & 0.5 & Fitted \\
Mass ($M_{\odot}$)  & 0.12 & Calculated \\
\hline
\multicolumn{4}{c}{Outer Wall of the Inner Gap} \\
\hline
Location (AU)  & 28 & Adopted/Refined & 4, 5\\
Height (AU) & 9 & Calculated \\
Illuminated Fraction (\%) & 66 & Calculated \\
Temperature of Illuminated Part (K) & 110 & Calculated \\
\hline
\multicolumn{4}{c}{Outer Wall of the Outer Gap} \\
\hline
Location (AU)  & 67 & Adopted/Refined & 4, 5\\
Height (AU) & 26 & Calculated \\
Illuminated Fraction (\%) & 10 & Calculated  \\
Temperature of Illuminated Part (K) & 70 & Calculated \\
\enddata
 \tablenotetext{a}{The mass accretion rate, viscosity coefficient, 
inclination, and grain properties are equal to those of the outer disk. 
The size is constrained by the 7 mm intensity upper limit towards the 
center.
}
 \tablerefs{(1) Manoj et al. 2006; (2) Dent et al. 2006; (3) Raman et
al. 2006; (4) Quanz et al. 2013; (5) This paper; (6) Grady et al. 2007.}

\end{deluxetable}

\end{document}